\documentclass[letterpaper]{article} % DO NOT CHANGE THIS
\usepackage{aaai2026}  % DO NOT CHANGE THIS
\usepackage{booktabs}   % \toprule \midrule \bottomrule
\usepackage{multirow}   % \multirow{...}{...}{...}
\usepackage{times}  % DO NOT CHANGE THIS
\usepackage{makecell}
\usepackage{helvet}  % DO NOT CHANGE THIS
\usepackage{courier}  % DO NOT CHANGE THIS
\usepackage[hyphens]{url}  % DO NOT CHANGE THIS
\usepackage{graphicx} % DO NOT CHANGE THIS
\def\UrlFont{\rm}  % DO NOT CHANGE THIS
\usepackage{natbib}  % DO NOT CHANGE THIS AND DO NOT ADD ANY OPTIONS TO IT
\usepackage{caption} % DO NOT CHANGE THIS AND DO NOT ADD ANY OPTIONS TO IT
\usepackage{algorithm}
\usepackage{algorithmic}
\usepackage{amsmath, amssymb}  % 引入 amssymb 即可支持 \mathbb

\usepackage{amsthm}

\newcommand{\KeyHeadword}[1]{\noindent\textbf{#1.}}
\newcommand{\firstpara}[1]{\noindent\textbf{{#1}.}~}

\usepackage{newfloat}
\usepackage{listings}
\DeclareCaptionStyle{ruled}{labelfont=normalfont,labelsep=colon,strut=off} % DO NOT CHANGE THIS
\floatstyle{ruled}
\newfloat{listing}{tb}{lst}{}
\floatname{listing}{Listing}
\title{Less Is More: Sparse and Cooperative Perturbation for Point Cloud Attacks}

\author{
Keke Tang\textsuperscript{\rm 1}\equalcontrib,
Tianyu Hao\textsuperscript{\rm 1}\equalcontrib,
Xiaofei Wang\textsuperscript{\rm 2}\equalcontrib, \\
Weilong Peng\textsuperscript{\rm 1}\thanks{W. Peng and P. Zhu are joint corresponding authors.},
Denghui Zhang\textsuperscript{\rm 1},
Peican Zhu\textsuperscript{\rm 3}\footnotemark[2],
Zhihong Tian\textsuperscript{\rm 1,4}
}

\affiliations{
\textsuperscript{\rm 1}Guangzhou University\\
\textsuperscript{\rm 2}University of Science and Technology of China\\
\textsuperscript{\rm 3}Northwestern Polytechnical University\\
\textsuperscript{\rm 4}Guangdong Key Laboratory of Industrial Control System Security\\
tangbohutbh@gmail.com,
howty666@gmail.com,
wxf9545@mail.ustc.edu.cn\\
wlpeng@tju.edu.cn,
denghui.zhang@gzhu.edu.cn,
ericcan@nwpu.edu.cn,
tianzhihong@gzhu.edu.cn
}

\begin{document}

\maketitle

\begin{abstract}

Most adversarial attacks on point clouds perturb a large number of points, causing widespread geometric changes and limiting applicability in real-world scenarios. While recent works explore sparse attacks by modifying only a few points, such approaches often struggle to maintain effectiveness due to the limited influence of individual perturbations. In this paper, we propose SCP, a sparse and cooperative perturbation framework that selects and leverages a compact subset of points whose joint perturbations produce amplified adversarial effects. Specifically, SCP identifies the subset where the misclassification loss is locally convex with respect to their joint perturbations, determined by checking the positive-definiteness of the corresponding Hessian block. The selected subset is then optimized to generate high-impact adversarial examples with minimal modifications. Extensive experiments show that SCP achieves 100\% attack success rates, surpassing state-of-the-art sparse attacks, and delivers superior imperceptibility to dense attacks with far fewer modifications.

%Most adversarial attacks on  point clouds perturb a large number of points, causing widespread geometric changes and limiting their applicability in real-world scenarios. While recent works explore sparse attacks by modifying only a few points, such approaches often struggle to maintain effectiveness due to the limited influence of individual perturbations. In this paper, we propose SCP, a sparse and cooperative perturbation framework that selects and leverages a compact subset of points whose joint perturbations produce amplified adversarial effects. Specifically, SCP identifies the subset where the misclassification loss is locally convex with respect to their joint perturbations, which is determined by checking the positive-definiteness of the corresponding Hessian block. The selected subset is then optimized to generate high-impact adversarial examples with minimal modifications. Extensive experiments show that SCP achieves 100\% attack success rates, surpassing state-of-the-art sparse attacks, and delivers superior imperceptibility to dense attacks with far fewer modifications. 

%Codes will be made public upon paper acceptance.

\if 0
Extensive experiments show that SCP significantly improves attack success rates over state-of-the-art sparse methods while achieving superior imperceptibility to dense attacks. Codes will be made public upon paper acceptance.
\fi

\end{abstract}

\section{Introduction}

With the growing adoption of 3D sensors and the advancement of deep learning techniques, deep neural networks (DNNs)~\cite{lecun2015deep} have become the dominant paradigm for point cloud perception~\cite{guo2020deep,ioannidou2017deep,Tang-DFN}, supporting applications such as autonomous driving and robotic manipulation.
However, recent studies have shown that these models are highly vulnerable to adversarial attacks, where subtle perturbations to input point clouds can cause severe misclassifications~\cite{Xiang-2019-3DAdversarialPCD}. 
This fragility has raised growing concerns and drawn increasing attention to adversarial attacks, which have become a valuable tool for exposing model weaknesses and promoting the development of more robust 3D perception systems.

%~\cite{sun2020adversarial,liu2021pointguard}.

\begin{figure}[!t]
\centering
\includegraphics[width=1.0\linewidth]{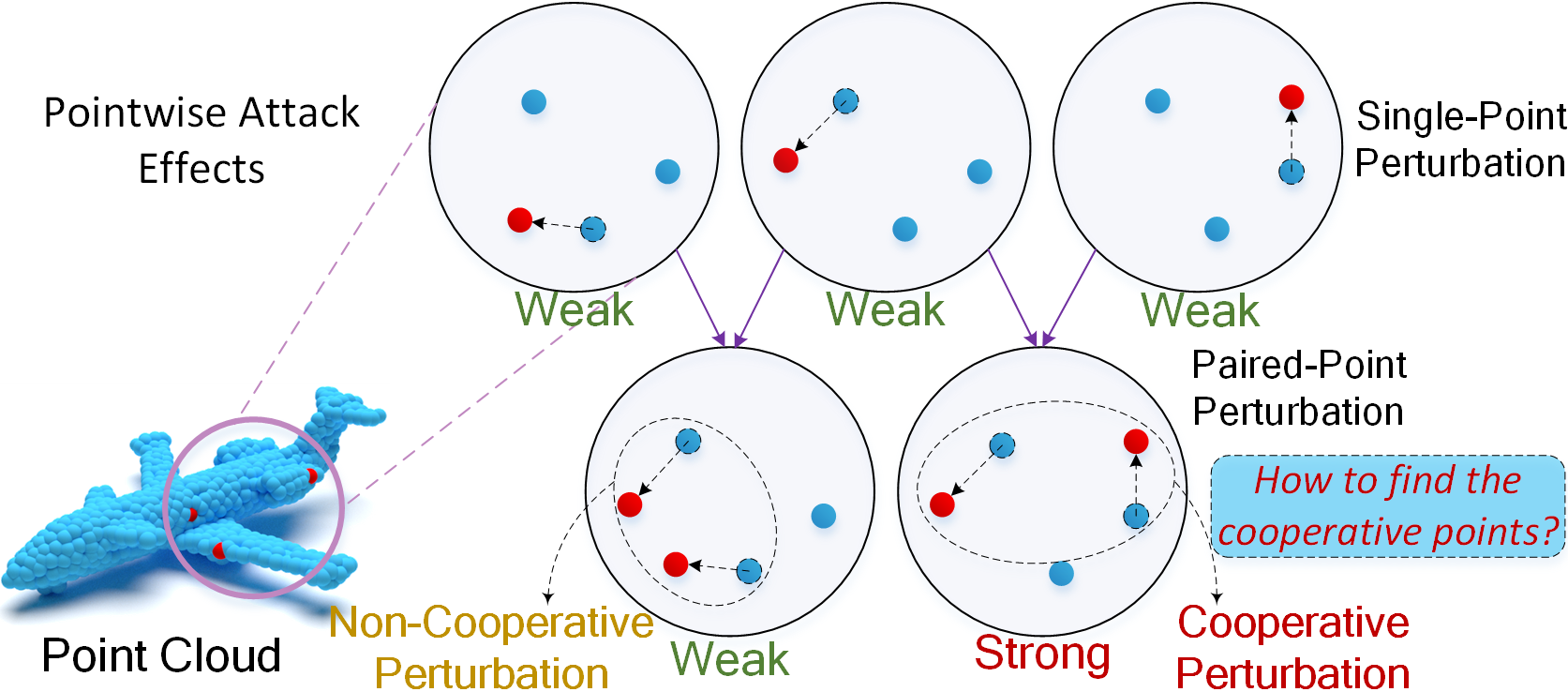} 

\caption{
Illustration of point-wise perturbation effects in point cloud attacks.
Single-point perturbations (top row) and non-cooperative combinations (bottom-left) often yield limited impact. In contrast, perturbing a cooperative subset (bottom-right) produces significantly stronger effects, motivating the need to identify such point groups.
}
\label{fig:teaser}

\end{figure}

\if 0
A large body of work has explored adversarial attacks on point clouds by applying point-wise perturbations across the entire input~\cite{Xiang-2019-3DAdversarialPCD,liu2022imperceptible,huang2022shape,tang2024manifoldConstraints,lou2024hide}. While effective, such dense perturbations can significantly distort the object's overall geometry, limiting their practicality in real-world applications. To address this, recent methods have proposed sparse attacks that modify only a small subset of points, using saliency cues~\cite{zheng2019pointcloud}, sparsity-constrained optimization~\cite{kim2021minimal,wicker2019robustness}, or semantic and geometric priors~\cite{shi2022shape}. Although these approaches reduce the number of modified points, they often rely on stronger per-point perturbations and may suffer degraded attack performance under distortion constraints.
\fi

A large body of work has explored adversarial attacks on point clouds by applying point-wise perturbations across the entire input~\cite{Xiang-2019-3DAdversarialPCD,liu2022imperceptible,huang2022shape,tang2024manifoldConstraints}. While effective, such dense perturbations can significantly distort the object's overall geometry, limiting their practicality in real-world applications. To address this, recent methods have proposed sparse attacks that modify only a small subset of points using sparsity-constrained optimization~\cite{kim2021minimal} or semantic and geometric priors~\cite{shi2022shape}. Although these methods reduce the number of modified points, they often require stronger per-point changes and suffer from degraded attack performance under distortion constraints.

A common assumption underlying these sparse methods is that point-wise contributions to the attack are independent. However, point cloud networks involve local aggregation and hierarchical encoding~\cite{Qi-2017-Pointnet,Qi-2017-Pointnet++}, which induce strong nonlinear interactions among points. As a result, perturbing certain points jointly can produce effects that are not captured when perturbing them individually. For example, two points within a semantically critical region may cooperatively activate or suppress key features, amplifying the adversarial effect, as illustrated in Fig.~\ref{fig:teaser}. We hypothesize that such cooperative subsets exist and that exploiting their joint influence enables stronger sparse attacks. This motivates us to move beyond isolated-point selection and instead identify compact subsets whose combined perturbation yields disproportionately strong adversarial impact.

In this paper, we propose SCP, a sparse and cooperative perturbation framework for point cloud attacks that perturbs a compact subset of points whose joint effect leads to strong adversarial behavior. Specifically, SCP first identifies a candidate set of influential points based on gradient information and then employs a greedy expansion strategy using Schur complement conditions to ensure the corresponding Hessian block is positive-definite, so that the constructed subset lies in a locally convex region of the loss landscape and thus exhibits cooperative influence that strengthens the adversarial effect.
The resulting subset is then jointly optimized to induce misclassification with minimal distortion. Extensive experiments show that SCP achieves 100\% attack success rates while modifying only a few points, significantly outperforming prior sparse attacks, and attains imperceptibility that is in most cases superior to dense attack methods.

Overall, our contribution is summarized as follows:

\begin{itemize}
\item We identify the overlooked role of cooperative interactions in sparse point cloud attacks and establish a Hessian-based criterion to characterize such synergy.
\item We develop SCP, a framework that selects compact cooperative subsets via gradient screening and Schur complement-guided expansion for joint perturbation.
\item We show by experiments that SCP achieves 100\% attack success with minimal modifications, outperforming  state-of-the-art sparse and dense attacks.

\end{itemize}

\section{Related Work}

\firstpara{Adversarial Attacks on 3D Point Clouds}
Adversarial attacks on point clouds can be broadly categorized into addition-based~\cite{Xiang-2019-3DAdversarialPCD}, deletion-based~\cite{zheng2019pointcloud,Yang-2019-AdvAttackAndDefense,zhang-2021-TopologyDestructionNetwork}, and perturbation-based methods~\cite{liu2022imperceptible}.
Among these, perturbation-based attacks~\cite{tang2022rethinking,tang2023deep,Tang-2024-FLAT,tang2024symattack,tang2025imperceptible,tang2025cage,tang2025MAT,tang2025P2S,wang2025eia} that modify the coordinates of existing points are the focus of this work.

Early efforts~\cite{Xiang-2019-3DAdversarialPCD,Liu-2019-extendingAdv3D} extended classic 2D attacks like FGSM~\cite{Goodfellow-2014-FGSM} and C\&W~\cite{Carlini-2017-cw} to the 3D domain. Subsequent work focused on improving stealthiness by preserving geometric structure, e.g., via curvature~\cite{wen2020geometry}, normal or tangent directions~\cite{liu2022imperceptible,huang2022shape}, or manifold priors~\cite{tang2024manifoldConstraints}. Generative approaches have also been explored through latent perturbation~\cite{Lee-2020-Shapeadv} and adversarial synthesis~\cite{Zhou-2020-LGGAN}.
Despite their success, these methods typically perturb most or all points, causing global geometric changes that limit real-world deployability and motivate the need for sparser alternatives.

\if 0
\firstpara{Sparse Adversarial Attacks on 3D Point Clouds}
To enhance stealthiness and real-world feasibility, several methods have sought to reduce the number of perturbed points in 3D adversarial attacks. Saliency-driven approaches identify high-impact points using gradient-based metrics; \citet{zheng2019pointcloud} construct point-wise saliency maps to guide perturbations toward critical regions. Optimization-based methods enforce sparsity through constrained formulations: \citet{wicker2019robustness} develop certified attacks with bounded perturbation regions, and \citet{kim2021minimal} jointly optimize point selection and distortion under cardinality and norm constraints. \citet{shi2022shape} further introduce shape priors to ensure semantic preservation during selective perturbation.
Despite their effectiveness, these methods typically assume independent point-wise influence and do not account for cooperative interactions. This often leads to higher per-point distortion or degraded performance under strict sparsity constraints. In contrast, our approach explicitly models second-order interactions to identify compact, mutually reinforcing subsets, enabling more effective and efficient sparse attacks.
\fi

\firstpara{Sparse Adversarial Attacks on 3D Point Clouds}
To improve stealthiness and enhance real-world applicability, sparse adversarial attacks restrict modifications to a limited subset of points. Existing methods can be broadly categorized into point removal/occlusion and point perturbation strategies.
Point removal or occlusion techniques achieve adversarial effects by deleting a small number of points. \citet{wicker2019robustness} removed either randomly selected points or those contributing most to max-pool features, while \citet{zheng2019pointcloud} eliminated points with the highest or lowest saliency scores to compromise shape integrity.
In contrast, perturbation-based methods manipulate a carefully selected set of points instead of removing them. \citet{kim2021minimal} jointly optimized point selection and displacement under $l_0$ constraints to minimize distortion, and \citet{shi2022shape} incorporated shape priors to preserve semantic consistency during selective perturbation.
Despite their effectiveness, these methods typically assume independent point-wise influence and do not account for cooperative interactions. This often leads to higher per-point distortion or degraded performance under strict sparsity constraints. In contrast, our approach explicitly models second-order interactions to identify compact, mutually reinforcing subsets, enabling more effective and efficient sparse attacks.

\if 0
\firstpara{Deep 3D Point Cloud Classification}
Recent advances in 3D point cloud classification have shifted from early voxel-based methods~\cite{Maturana-2015-voxnet} to direct point set processing frameworks such as PointNet~\cite{Qi-2017-Pointnet} and PointNet++~\cite{Qi-2017-Pointnet++}. Subsequent developments include point-based convolutional networks~\cite{Wu-2019-Pointconv,Li-2018-PointCNN}, graph-based architectures~\cite{Wang-2019-DGCNN,Shi-2020-PointGNN}, and transformer-based models~\cite{PT,PT2,PT3}, which capture local geometric structure and global context more effectively.
We aim to attack these classifiers in a sparse manner.
\fi

\firstpara{Deep 3D Point Cloud Classification}
Modern 3D point cloud classifiers typically operate directly on point sets, beginning with PointNet~\cite{Qi-2017-Pointnet} and its hierarchical variant PointNet++~\cite{Qi-2017-Pointnet++}. Later advancements include point-based convolutional networks~\cite{Wu-2019-Pointconv}, graph-based architectures~\cite{Wang-2019-DGCNN}, and transformer-based models~\cite{PT,PT2}, which better capture local geometry and global context. We aim to attack these classifiers in a sparse manner.

\section{Problem Formulation}

\subsection{Problem Statement of Sparse Adversarial Attacks}

\firstpara{Preliminary on Adversarial Attacks} 
Given a point cloud $\mathcal{P} \in \mathbb{R}^{n \times 3}$ with its corresponding label $y \in \{1, \dots, C\}$, where $C$ is the number of semantic categories, the goal of perturbation-based adversarial attacks is to mislead a 3D deep classification model $\mathcal{F}$ into making incorrect predictions. This is achieved by crafting an adversarial point cloud $\mathcal{P}^{adv}$ through the application of imperceptible perturbations to the original points.

Formally, the adversarial point cloud is defined as:
\begin{equation}
\mathcal{P}^{adv} = \mathcal{P} + \boldsymbol{\Delta} = \mathcal{P} + 
\begin{bmatrix}
\Delta P_1^\top & \cdots & \Delta P_n^\top
\end{bmatrix}^\top,
\end{equation}
where $\boldsymbol{\Delta} \in \mathbb{R}^{n \times 3}$ is the perturbation matrix, and each $\Delta P_i \in \mathbb{R}^{1 \times 3}$ is a row vector perturbing point $P_i$.

\if 0
Formally, the adversarial point cloud is defined as:
\begin{equation}
\mathcal{P}^{adv} = \mathcal{P} + \boldsymbol{\Delta} = \mathcal{P} + 
\begin{bmatrix}
\Delta P_1 \\
\vdots \\
\Delta P_n
\end{bmatrix},
\end{equation}
where $\boldsymbol{\Delta} \in \mathbb{R}^{n \times 3}$ is the perturbation matrix, and each $\Delta P_i \in \mathbb{R}^{1 \times 3}$ is a row vector representing the perturbation applied to point $P_i$.
\fi

\begin{figure*}[!t]
    \centering
    \includegraphics[width=0.73\linewidth]{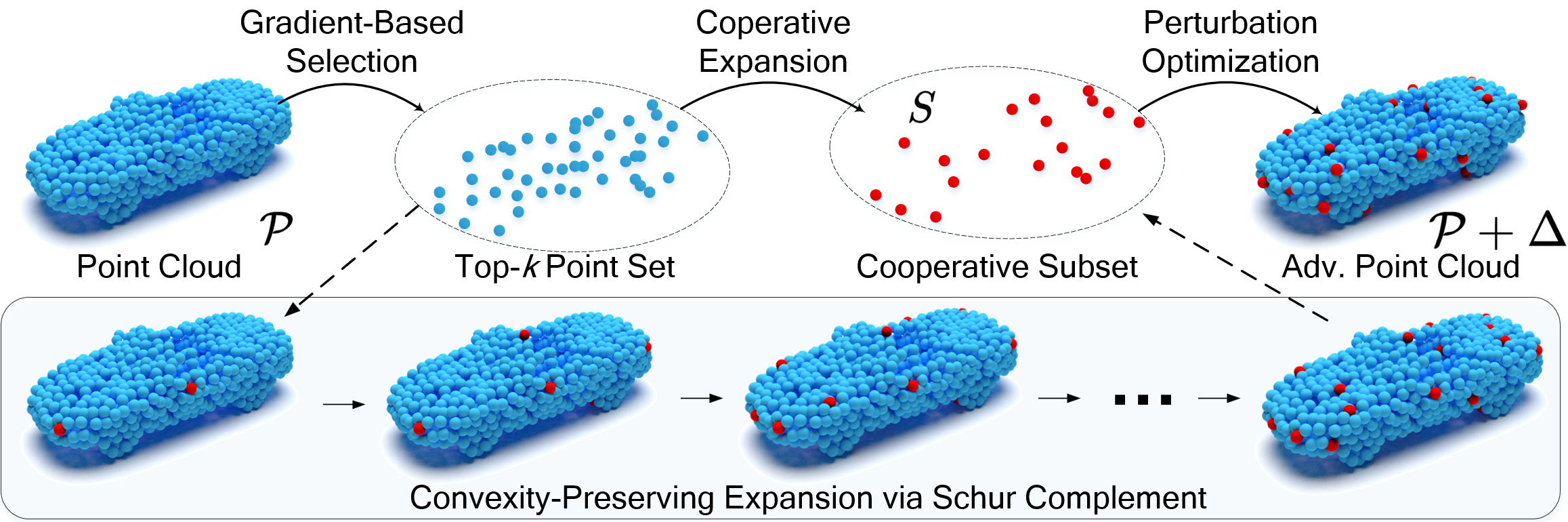}

    \caption{
    Overview of the SCP framework.
   Starting from a point cloud \(\mathcal{P}\), we first apply gradient-based selection to extract a top-\(k\) point set. 
  Then, convexity-preserving expansion via Schur complement yields a cooperative subset. 
  Finally, perturbation optimization is performed on this subset to obtain the perturbation \(\Delta\).    }
    \label{fig:method}
\end{figure*}

The perturbation is typically obtained by solving the following optimization problem:
\begin{equation}
\min_{\boldsymbol{\Delta}} \; 
L_{mis}(\mathcal{F}, \mathcal{P} + \boldsymbol{\Delta}, y) + \beta_1 D(\mathcal{P}, \mathcal{P} + \boldsymbol{\Delta}),
\label{eq:formal}
\end{equation}
where $L_{mis}(\cdot, \cdot, \cdot)$ is a misclassification loss (e.g., the negative cross-entropy), $D(\cdot, \cdot)$ measures the distortion to enforce imperceptibility, and $\beta_1$ balances the two objectives. While this formulation is typically used for untargeted attacks, it can also be readily adapted for targeted scenarios.

\firstpara{Sparse Adversarial Attacks}
Sparse adversarial attacks aim to perturb only a small subset of points in the input point cloud, rather than modifying all points. Such selective perturbations are considered more stealthy and physically realizable, especially when large-scale modifications are impractical or easily detectable.

Let $S = \{s_1, s_2, \dots, s_t\}$ denote the indices of perturbed points, and let $\{\Delta P_{s_i}\}_{i=1}^t$ be their corresponding perturbations. The adversarial point cloud is then given by:
\begin{equation}
\mathcal{P}^{adv} = \mathcal{P} + \sum_{i=1}^t \left( \mathbf{e}_{s_i} \otimes \Delta P_{s_i} \right),
\end{equation}
where $\mathbf{e}_{s_i} \in \mathbb{R}^{n}$ is a standard basis vector with a 1 in the $s_i$-th position and zeros elsewhere, and $\otimes$ denotes the Kronecker product. Each term yields a matrix in $\mathbb{R}^{n \times 3}$ whose $s_i$-th row equals $\Delta P_{s_i}$ and all other rows are zero.

\KeyHeadword{Discussion}
Sparse adversarial attacks typically modify only a small subset of points using sparsity-constrained optimization~\cite{kim2021minimal} or by incorporating semantic and geometric priors~\cite{shi2022shape}. While these approaches reduce the number of modified points, they generally assume independent perturbation effects and overlook interactions among the selected points. As a result, sparse attacks commonly require stronger per-point distortions to remain effective, which may degrade their imperceptibility or lead to reduced attack success under strict distortion budgets.

\subsection{Cooperative Behavior in Sparse Perturbations}

While sparse adversarial attacks aim to perturb only a limited number of points, the overall adversarial effectiveness depends not only on the strength of individual perturbations but also on how these perturbations interact. In particular, a small set of perturbations can produce a disproportionately large impact if they exhibit \emph{cooperative behavior}, where their joint effect exceeds the sum of their individual effects. We formalize this notion below.

\firstpara{Definition of Cooperation via Jensen's Inequality}
Let $\{\Delta P_{s_1}, \dots, \Delta P_{s_t}\}$ be the perturbations applied to selected points $\{s_1, \dots, s_t\}$. We say these perturbations are \emph{cooperative} if the following strict Jensen-type inequality holds:
\begin{align}
&L_{mis}\left(\mathcal{F}, \mathcal{P} + \sum_{i=1}^t \alpha_i (\mathbf{e}_{s_i} \otimes \Delta P_{s_i}), y \right) \nonumber \\
&\quad > \sum_{i=1}^{t} \alpha_i \cdot 
L_{mis}\left(\mathcal{F}, \mathcal{P} + \mathbf{e}_{s_i} \otimes \Delta P_{s_i}, y \right),
\label{eq:cooperation}
\end{align}
where $\alpha_i > 0$ and $\sum_{i=1}^t \alpha_i = 1$. This inequality implies that the joint loss induced by the weighted combination of perturbations exceeds the weighted average of individual losses, indicating synergy among the selected perturbations.

\firstpara{A Sufficient Condition for Identifying Cooperation}
To characterize when cooperative effects arise, we consider the local curvature of the misclassification loss. Define the concatenated perturbation vector as
\begin{equation}
\boldsymbol{\delta}_{S} = 
\begin{bmatrix}
\Delta P_{s_1} & \Delta P_{s_2} & \cdots & \Delta P_{s_t}
\end{bmatrix}^\top \in \mathbb{R}^{3t \times 1}.
\end{equation}
A sufficient condition for the strict inequality in Eq.~\eqref{eq:cooperation} to hold is that the loss function \(L_{mis}\) is locally strictly convex with respect to \(\boldsymbol{\delta}_S\)~\cite{boyd2004convex}, which requires that the Hessian matrix
\begin{equation}
\mathbf{H}(\boldsymbol{\delta}_S) = \nabla^2_{\boldsymbol{\delta}_{S}} L_{mis}\left(\mathcal{F}, \mathcal{P} + \sum_{i=1}^t \alpha_i (\mathbf{e}_{s_i} \otimes \Delta P_{s_i}), y\right)
\end{equation}
is positive definite, i.e.,
\begin{equation}
\mathbf{H}(\boldsymbol{\delta}_S) \succ 0.
\end{equation}
This ensures that the perturbation combination lies in a region of positive curvature, resulting in a joint adversarial effect greater than the sum of individual effects.

This condition underpins our approach for identifying cooperative perturbation subsets and guides the selection of sparse yet effective adversarial points.

\section{Method}

In this section, we present SCP, a sparse and cooperative perturbation framework for point cloud attacks. The framework consists of two stages: we first identify a subset of points whose perturbations exhibit cooperative behavior, and then optimize perturbations over this subset to generate effective adversarial point clouds. An overview of the pipeline is illustrated in Fig.~\ref{fig:method}.

\subsection{Selection of Cooperative Subset}

Directly computing the full Hessian matrix of the misclassification loss with respect to all point-wise perturbations, and subsequently identifying its largest positive definite submatrix, is computationally infeasible.  
To address this, SCP adopts a three-stage greedy selection strategy:
(i) identifying influential points via gradient analysis,
(ii) constructing a convex cooperative subset via Schur complement checks,
and (iii) refining the subset with relaxed inclusion criteria.

\firstpara{Gradient-Based Selection of Influential Points}
We first compute the gradient of the misclassification loss with respect to each point in the input point cloud:
\begin{equation}
    g_i = \left\| \nabla_{P_i} L_{mis}(\mathcal{F}, \mathcal{P}, y) \right\|_2.
\end{equation}
Points with larger gradient magnitudes indicate greater local influence on the loss. We thus select the top-$k$ points ranked by $g_i$ as the initial candidate set for cooperative expansion.

\firstpara{Convexity-Preserving Expansion via Schur Complement}
Given an initial cooperative set, we iteratively evaluate candidate points to determine whether their inclusion preserves the local positive definiteness of the associated Hessian block. 

Let the current cooperative set correspond to a Hessian submatrix \(\mathbf{A}\), a candidate point contribute a block \(\mathbf{C}\), and the interaction terms be represented by a coupling block \(\mathbf{B}\). The augmented Hessian has the structure:
\begin{equation}
\mathbf{H}^{'} =
\begin{bmatrix}
\mathbf{A} & \mathbf{B} \\
\mathbf{B}^\top & \mathbf{C}
\end{bmatrix}.
\end{equation}
By the Schur complement condition, the augmented matrix \(\mathbf{H}^{'}\) is positive definite if and only if \(\mathbf{A} \succ 0\) and
\begin{equation}
\mathbf{C} - \mathbf{B}^\top \mathbf{A}^{-1} \mathbf{B} \succ 0.
\end{equation}
We define the \emph{Schur Surplus} as the minimum eigenvalue of the residual block:
\begin{equation}
    S_s = \lambda_{\min}\left( \mathbf{C} - \mathbf{B}^\top \mathbf{A}^{-1} \mathbf{B} \right).
\end{equation}
If \(S_s > 0\), the candidate point is deemed cooperative and added to the current subset.

\firstpara{Tolerance-Based Inclusion of Marginal Points}
To improve flexibility and account for numerical imprecision, we relax the strict Schur condition by introducing a tolerance threshold \(\epsilon > 0\). A candidate point is accepted if the corresponding Schur Surplus satisfies
\begin{equation}
S_s > -\epsilon.
\end{equation}
This  relaxation permits inclusion of marginal candidates that may still yield cooperative effects, even if the strict positive definiteness condition is not fully met.

\if 0
\subsection{Perturbation Optimization on Cooperative Subset}

Given the selected cooperative subset \( S = \{s_1, \dots, s_t\} \), we optimize the associated perturbations \( \{\Delta P_{s_i}\} \) to generate the adversarial example. The objective is formulated as:
\begin{equation}
\min_{\{\Delta P_{s_i}\}} \; L_{mis}\left(\mathcal{F}, \mathcal{P} + \sum_{i=1}^t \left( \mathbf{e}_{s_i} \otimes \Delta P_{s_i} \right), y \right) + \beta_1 D,
\label{eq:scopt}
\end{equation}
where \( D(\cdot,\cdot) \) denotes the distortion penalty, e.g., based on Chamfer distance and Hausdorff distance, and \( \beta_1 \) balances attack success and imperceptibility.
\fi

\subsection{Perturbation Optimization on Cooperative Subset}  
Given the selected cooperative subset \( S = \{s_1, \dots, s_t\} \), we optimize the associated perturbations \( \{\Delta P_{s_i}\} \) by specializing the general objective in Eqn.~\ref{eq:formal} to this subset, which leads to the following formulation:
\begin{equation}
\min_{\{\Delta P_{s_i}\}} \; 
L_{mis}\left(\mathcal{F}, \mathcal{P} + \sum_{i=1}^t \left( \mathbf{e}_{s_i} \otimes \Delta P_{s_i} \right), y \right) + \beta_1 D,
\label{eq:scopt}
\end{equation}
where \( \beta_1 D \) is the distortion penalty term defined in Eqn.~\ref{eq:formal}.

The optimization is conducted following \cite{Xiang-2019-3DAdversarialPCD}.

\if 0
The optimization is conducted %using projected gradient descent, 
%following the setting in
%~
following \cite{Xiang-2019-3DAdversarialPCD}, and continues until either successful misclassification is achieved or the maximum number of iterations is reached.
\fi

\section{Experiments}

\subsection{Experimental Setup}

\firstpara{Implementation}  
We implement SCP in PyTorch~\cite{Pytorch}. The cooperative subset selection starts from $k = 256$ candidate points, and the tolerance parameter is set to $\epsilon = 10^{-6}$. Perturbation optimization is performed using the Adam optimizer ($\beta_1 = 0.9$, $\beta_2 = 0.999$) with a learning rate of $0.01$, running $10$ binary search rounds and $500$ gradient steps. All experiments are conducted on a workstation with dual 2.40 GHz CPUs,  and eight NVIDIA RTX 3090 GPUs.

\firstpara{Datasets}  
We evaluate our attack on ModelNet40~\cite{wu20153d} and  ScanObjectNN~\cite{uy2019-ScanObjectNN}. All point clouds are uniformly resampled to 1,024 points.

%~\cite{Xiang-2019-3DAdversarialPCD}. 

%Results on ShapeNet Part are provided in the supplementary material.

\if 0
\firstpara{Victim 3D DNN Classifiers}  
We assess the performance of our attack using four widely adopted 3D point cloud classifiers with diverse architectures: PointNet~\cite{Qi-2017-Pointnet}, DGCNN~\cite{Wang-2019-DGCNN}, Point Transformer (PTv1)~\cite{PT}, and the recent Mamba3D~\cite{liang2024pointmamba}. All models are trained according to the settings described in their original implementations.
\fi

\if 0
\firstpara{Victim 3D DNN Classifiers}  
We evaluate our attack on four widely used 3D point cloud classifiers with diverse architectures: PointNet~\cite{Qi-2017-Pointnet}, DGCNN~\cite{Wang-2019-DGCNN}, Point Transformer (PTv1)~\cite{PT}, and Mamba3D~\cite{liang2024pointmamba}. All models follow the training settings of their original implementations.
\fi

\firstpara{Victim Models}  
We evaluate our attack on four DNN  classifiers: PointNet~\cite{Qi-2017-Pointnet}, DGCNN~\cite{Wang-2019-DGCNN}, PTv1~\cite{PT}, and Mamba3D~\cite{liang2024pointmamba}, using their original training settings.

\begin{table*}[!t]
\centering

\setlength{\tabcolsep}{2.2pt}  % tighter column spacing
\scalebox{0.81}{
\begin{tabular}{ll*{7}{c}@{\hspace{4pt}}*{7}{c}}
\toprule
\multicolumn{2}{c}{} &
\multicolumn{7}{c}{{ModelNet40}} &
\multicolumn{7}{c}{{ScanObjectNN}}\\
\cmidrule(lr){3-9}\cmidrule(lr){10-16}
{Model} & {Metric} &
AdvStick & RandOcc & CritOcc & HighSal & LowSal & MiniPert & Ours &
AdvStick & RandOcc & CritOcc & HighSal & LowSal & MiniPert & Ours\\
\midrule
\multirow{4}{*}{\rotatebox{90}{PointNet}}
 & ASR            & 83.70 & 53.23 & 21.74 & 60.47 & 57.80 & 89.38 & \textbf{100} &
                     87.50 & 60.19 & 43.83 & 67.12 & 64.99 & 91.72 & \textbf{100}\\
 & CD ($10^{-4}$) & 49.30 &  6.94 &  1.27 &  8.41 &  6.82 &  1.55 & \textbf{0.74} &
                     51.80 &  5.71 &  1.80 &  6.73 &  5.87 &  1.12 & \textbf{0.93}\\
 & HD ($10^{-2}$) & 14.90 & \textbf{0.25} & 1.03 & \textbf{0.25} & \textbf{0.25} & 1.88 & 0.32 &
                      16.7 & 0.25 & 2.57 & \textbf{0.24} & 0.25 & 1.15 & 0.42\\
 & \# Points      & 210   & 413   &  50  & 200  & 200  &  36  & \textbf{34} &
                     210   & 397   &  70  & 200  & 200  &  34  & \textbf{32}\\
\midrule
\multirow{4}{*}{\rotatebox{90}{DGCNN}}
 & ASR            & 73.70 & 40.20 &  8.87 & 46.22 & 43.08 & 73.64 & \textbf{100} &
                     85.90 & 44.90 & 28.64 & 38.02 & 33.92 & 78.29 & \textbf{100}\\
 & CD ($10^{-4}$) & 17.54 &  7.68 &  1.31 & 10.47 & 11.49 &  5.10 & \textbf{1.06} &
                     10.45 &  7.03 &  2.38 &  7.83 &  6.33 &  2.67 & \textbf{1.68}\\
 & HD ($10^{-2}$) &  4.68 & \textbf{0.24} & 1.05 & 0.26 & 0.25 & 1.91 & 0.84 &
                      4.01 & \textbf{0.23} & 2.34 & 0.25 & 0.28 & 1.15 & 0.46\\
 & \# Points      & 500   & 689   &  86  & 200  & 200  & 138  & \textbf{62} &
                     500   & 712   &  93  & 200  & 200  & 110  & \textbf{60}\\
\midrule
\multirow{4}{*}{\rotatebox{90}{PTv1}}
 & ASR            & 61.58 & 42.76 & 12.83 & 37.25 & 32.10 & 54.69 & \textbf{100} &
                     74.70 & 61.07 & 25.63 & 31.56 & 27.71 & 49.22 & \textbf{100}\\
 & CD ($10^{-4}$) & 16.79 &  6.37 &  1.36 & 13.24 & 11.25 &  6.78 & \textbf{1.01} &
                      8.10 &  6.51 &  2.57 &  6.02 &  5.37 &  3.70 & \textbf{0.42}\\
 & HD ($10^{-2}$) &  5.30 &  0.26 &  1.16 &  \textbf{0.25} &  \textbf{0.25} &  2.50 & 0.88 &
                      6.13 &  0.25 &  2.21 &  \textbf{0.24} &  0.25 &  0.86 & 0.78\\
 & \# Points      & 500   & 427   &  114  & 200  & 200  & 207  & \textbf{52} &
                     500   & 488   &  125  & 200  & 200  & 206  & \textbf{53}\\
\midrule
\multirow{4}{*}{\rotatebox{90}{Mamba3D}}
 & ASR            & 74.91 & 50.64 &  7.51 & 51.06 & 48.65 & 73.32 & \textbf{100} &
                     88.15 & 65.15 & 22.46 & 53.09 & 49.36 & 72.45 & \textbf{100}\\
 & CD ($10^{-4}$) & 17.57 &  9.29 &  1.41 & 9.52 & 10.76 &  2.27 & \textbf{0.83} &
                      8.35 &  8.12 &  2.70 & 8.79 &  8.22 &  1.11 & \textbf{0.68}\\
 & HD ($10^{-2}$) &  4.33 &  0.26 &  1.15 &  \textbf{0.25} & 0.26 &  0.88 & 0.79 &
                      3.98 &  \textbf{0.24} &  2.70 &  0.25 &  0.25 &  1.09 & 0.90\\
 & \# Points      & 500   & 482   &  97  & 200  & 200  &  74  & \textbf{49} &
                     500   & 473   &  106  & 200  & 200  &  52  & \textbf{51}\\
\bottomrule
\end{tabular}}
\caption{Comparison on ASR, imperceptibility (CD and HD), and the number of modified  points (\# Points) for different sparse attacks across four DNN classifiers on {ModelNet40} and {ScanObjectNN}.}
\label{tab:sparseAttack}
\end{table*}

\begin{table*}[!t]
\setlength{\tabcolsep}{3pt}  % tighter column spacing
\centering
%\caption{Comparison on the perturbation sizes required by different methods to reach their highest achievable ASR. The evaluation is conducted across different DNN classifiers on  ModelNet40 and ScanObjectNN.}

\setlength{\tabcolsep}{3.6pt}  % tighter column spacing
\scalebox{0.81}{%
  \begin{tabular}{>{\centering\arraybackslash}p{1.55cm}lcccccccccccccccc}
  \toprule
  \multicolumn{2}{c}{} & \multicolumn{8}{c}{{ModelNet40}} & \multicolumn{8}{c}{{ScanObjectNN}}\\
  \cmidrule(lr){3-10}\cmidrule(lr){11-18}
  \multicolumn{1}{c}{{Model}} & \multicolumn{1}{c}{{Attack}} &
    ASR & CD & HD & $l_2$ & GR & Curv & EMD & \# Points &
    ASR & CD & HD & $l_2$ & GR & Curv & EMD & \# Points \\
    & & (\%) & ($10^{-4}$) & ($10^{-2}$) &  &  & ($10^{-2}$) & ($10^{-2}$) & & (\%) & ($10^{-4}$) & ($10^{-2}$) &  &  & ($10^{-2}$) & ($10^{-2}$) \\
  \midrule
  % ---------- PointNet ----------
  \multirow{7}{*}{\rotatebox{90}{{PointNet}}}
   & PGD      & 100 & 26.582 & 10.777 & 2.804 & 0.406 & 5.681 & 5.705 & 1003 &
                 100 & 20.12 & 7.314 & 2.722 & 0.307 & 7.641 & 4.993 & 1019 \\
   & IFGM     & 100 &  8.982 &  9.859 & 1.195 & 0.368 & 1.209 & 1.537 &  670 &
                 100 &  5.915 & 6.297 & 0.993 & 0.258 & 1.830 & 1.463 &  762 \\
   & GeoA$^{3}$ & 100 &  4.869 &  0.524 & 1.423 & 0.215 & 0.348 & 2.415 &  957 &
                 100 &  3.425 & 0.655 & 1.355 & 0.128 & 0.574 & 0.493 &  933 \\
   & AOF      & 100 & 18.063 &  1.204 & 1.896 & 0.161 & 3.719 & 3.575 &  966 &
                 100 &  5.842 & 0.484 & 0.924 & 0.159 & 3.455 & 1.814 &  675 \\
   & SI-Adv   & 100 &  2.855 &  2.259 & 0.779 & 0.183 & 0.276 & 0.783 &  998 &
                 100 &  1.120 & 1.122 & \textbf{0.400} & 0.116 & 2.154 & 0.493 &  972 \\
   & {Ours} & 100 &  \textbf{0.741} &  \textbf{0.315} &  \textbf{0.521} & \textbf{0.160} &  \textbf{0.148} &  \textbf{0.199} &  \textbf{34} &
                 100 &  \textbf{0.931} &  \textbf{0.416} & 0.665 & \textbf{0.091} &  \textbf{0.133} &  \textbf{0.167} &  \textbf{32} \\
  \midrule
  % ---------- DGCNN ----------
  \multirow{7}{*}{\rotatebox{90}{{DGCNN}}}
   & PGD      & 100 & 28.517 &  8.772 & 2.853 & 0.347 & 5.915 & 5.927 & 1024 &
                 100 & 19.29 & 1.030 & 2.766 & 0.145 & 7.963 & 5.091 & 1024 \\
   & IFGM     & 100 & 10.701 &  7.148 & 1.622 & 0.363 & 2.849 & 3.777 & 1024 &
                 100 &  4.821 & 0.842 & 0.891 & 0.120 & 3.349 & 2.256 & 1024 \\
   & GeoA$^{3}$ & 100 &  13.017 &  2.059 & 1.589 & 0.169 & 1.700 & 2.279 & 1024 &
                 100 &  9.328 & 0.886 & 1.508 & 0.213 & 4.337 & 2.241 & 1024 \\
   & AOF      & 100 & 25.229 &  1.249 & 2.351 & 0.161 & 4.325 & 4.959 & 1024 &
                 100 &  7.973 & \textbf{0.375} & 1.084 & 0.107 & 4.262 & 2.513 & 1024 \\
   & SI-Adv   & 100 &  8.911 &  1.811 & \textbf{1.348} & \textbf{0.136} & 1.317 & 2.908 & 1024 &
                 100 &  3.007 & 0.555 & 0.752 & \textbf{0.096} & 1.276 & 1.526 & 1024 \\
   & {Ours} & 100 &  \textbf{1.059} &  \textbf{0.841} & 1.378 & 0.150 &  \textbf{0.385} &  \textbf{0.719} &  \textbf{62} &
                 100 &  \textbf{1.675} & 0.458 &  \textbf{0.691} & 0.163 &  \textbf{0.383} &  \textbf{0.384} &  \textbf{60} \\
  \midrule
  % ---------- PointTran ----------
  \multirow{7}{*}{\rotatebox{90}{{PTv1}}}
   & PGD      & 100 & 24.521 & 1.724 & 2.584 & 0.181 & 5.548 & 5.589 & 1024 &
                 100 & 17.79 & 1.484 & 2.636 & 0.136 & 7.483 & 4.899 & 1024 \\
   & IFGM     & 100 &  7.544 & 5.563 & 1.082 & 0.520 & 0.900 & 1.704 & 1024 &
                 100 &  2.778 & 0.464 & 0.819 & 0.107 & 2.278 & 1.592 & 1024 \\
   & GeoA$^{3}$ & 100 &  7.815 & 3.296 & 0.973 & 0.199 & 1.618 & 2.357 & 1024 &
                 100 &  6.452 & 0.830 & 1.050 & 0.122 & 4.259 & 2.132 & 1024 \\
   & AOF      & 100 & 31.840 & 1.164 & 3.065 & 0.194 & 6.943 & 5.581 & 1024 &
                 100 &  11.96 & \textbf{0.370} & 1.029 & 0.106 & 4.220 & 2.432 & 1024 \\
   & SI-Adv   & 100 & 13.456 & 3.044 & 1.806 & 0.186 & 2.334 & 3.385 & 1024 &
                 100 &  6.833 & 1.186 & 1.221 & 0.115 & 2.253 & 2.090 & 1024 \\
   & {Ours} & 100 &  \textbf{1.010} &  \textbf{0.877} &  \textbf{0.795} & \textbf{0.139} &  \textbf{0.168} &  \textbf{0.314} &  \textbf{52} &
                 100 &  \textbf{0.415} & 0.784 &  \textbf{0.366} & \textbf{0.106} &  \textbf{0.250} &  \textbf{0.183} &  \textbf{53} \\
  \midrule
  % ---------- Mamba3D ----------
  \multirow{7}{*}{\rotatebox{90}{{Mamba3D}}}
   & PGD      & 100 & 28.407 & 3.170 & 2.763 & 0.229 & 6.084 & 5.872 & 1024 &
                 100 & 21.27 & 1.371 & 2.604 & 0.159 & 8.108 & 5.162 & 1024 \\
   & IFGM     & 100 &  6.075 & 1.652 & \textbf{0.854} & 0.179 & 1.827 & 1.979 & 1024 &
                 100 &  4.902 & 1.521 & 0.962 & 0.146 & 3.211 & 1.935 & 1024 \\
   & GeoA$^{3}$ & 100 &  8.833 & 1.772 & 1.019 & 0.255 & 1.959 & 2.043 & 1024 &
                 100 &  5.748 & 0.899 & 1.083 & 0.177 & 3.588 & 2.014 & 1024 \\
   & AOF      & 100 & 18.073 & \textbf{0.711} & 1.619 & \textbf{0.137} & 3.413 & 4.033 & 1024 &
                 100 &  8.238 & \textbf{0.262} & 0.967 & \textbf{0.103} & 3.983 & 2.505 & 1024 \\
   & SI-Adv   & 100 &  5.685 & 1.722 & 0.965 & 0.152 & 0.617 & 1.940 & 1024 &
                 100 &  2.836 & 0.887 & \textbf{0.868} & 0.127 & 0.833 & 1.256 & 1024 \\
   & {Ours} & 100 &  \textbf{0.831} & 0.794 & 1.010 & 0.158 &  \textbf{0.514} &  \textbf{0.581} &  \textbf{49} &
                 100 &  \textbf{0.682} & 0.904 & 0.879 & 0.138 &  \textbf{0.294} &  \textbf{0.271} &  \textbf{51} \\
  \bottomrule
  \end{tabular}%
}
\caption{
Comparison on the perturbation sizes required by dense attack methods and our sparse SCP to reach their highest achievable ASR, evaluated across different DNN classifiers on ModelNet40 and ScanObjectNN.
}
\label{tab:dense_impercep_metrics}
\end{table*}

\if 0
\firstpara{Baselines}  
For sparse attacks, we include \textit{Adversarial Stick (AdvStick)}~\cite{liu2020adversarial}, which inserts stick-like structures outside the point cloud; \textit{Random Occlusion (RandOcc)} and \textit{Critical Occlusion (CritOcc)}~\cite{wicker2019robustness}, which remove random points or those contributing most to max-pool features; \textit{High Saliency (HighSal)} and \textit{Low Saliency (LowSal)}~\cite{zheng2019pointcloud}, which drop points with the highest or lowest saliency scores; and \textit{Minimal Perturbation (MiniPert)}~\cite{kim2021minimal}, which jointly optimizes point selection and displacement under $l_0$ regularization.  
For dense attacks, we select the gradient-based \textit{PGD} and \textit{IFGM}~\cite{Dong-2020-FGM_PGD_IMPLEMENT}, the geometry-aware optimization method \textit{GeoA$^3$}~\cite{wen2020geometry}, the frequency-domain approach \textit{AOF}~\cite{liu2022-AOF}, and the direction-guided \textit{SI-Adv}~\cite{huang2022shape}.
\fi

\firstpara{Baseline Methods}  
For sparse attacks, we include \textit{Adversarial Stick (AdvStick)}~\cite{liu2020adversarial}, which inserts stick-like structures; \textit{Random Occlusion (RandOcc)} and \textit{Critical Occlusion (CritOcc)}~\cite{wicker2019robustness}, which remove random points or those contributing most to max-pool features; \textit{High Saliency (HighSal)} and \textit{Low Saliency (LowSal)}~\cite{zheng2019pointcloud}, which drop points with the highest or lowest saliency scores; and \textit{Minimal Perturbation (MiniPert)}~\cite{kim2021minimal}, which jointly optimizes point selection and displacement under $l_0$ regularization.  
For dense attacks, we select the gradient-based \textit{PGD} and \textit{IFGM}~\cite{Dong-2020-FGM_PGD_IMPLEMENT}, the geometry-aware optimization method \textit{GeoA$^3$}~\cite{wen2020geometry}, the frequency-domain approach \textit{AOF}~\cite{liu2022-AOF}, and the direction-guided \textit{SI-Adv}~\cite{huang2022shape}.

\firstpara{Evaluation Setting and Metrics}  
For sparse attack methods, we follow the configurations reported in their original papers and evaluate performance using attack success rate (ASR), Chamfer distance (CD)~\cite{fan2017point}, Hausdorff distance (HD)~\cite{taha2015metrics}, and the number of modified points (\# Points) as an indicator of sparsity.
For dense attacks and our SCP framework, we configure each approach to achieve its maximum ASR. Under this maximal adversarialness condition~\cite{tang2024manifoldConstraints}, we further evaluate imperceptibility using additional metrics: $l_2$-norm ($l_2$), Curvature (Curv), Geometric Regularity (GR)~\cite{wen2020geometry}, and Earth Mover’s Distance (EMD)~\cite{rubner2000earth}.

\subsection{Comparison and Performance Analysis}

\firstpara{Comparison with Sparse Attacks}  
We first compare SCP with representative sparse attack methods. As shown in Tab.~\ref{tab:sparseAttack}, SCP consistently achieves the highest attack success rates across all datasets and classifiers, reaching 100\% in all cases. In contrast, the success rates of other methods remain far below this level, often failing to surpass 80\%. Furthermore, these baselines produce significantly larger CD and HD values, indicating lower imperceptibility. They also tend to modify more points, whereas SCP attains stronger attacks with sparser changes to the input point clouds.

  \begin{figure*}[!t]
    \centering
    \includegraphics[width=0.88\linewidth]{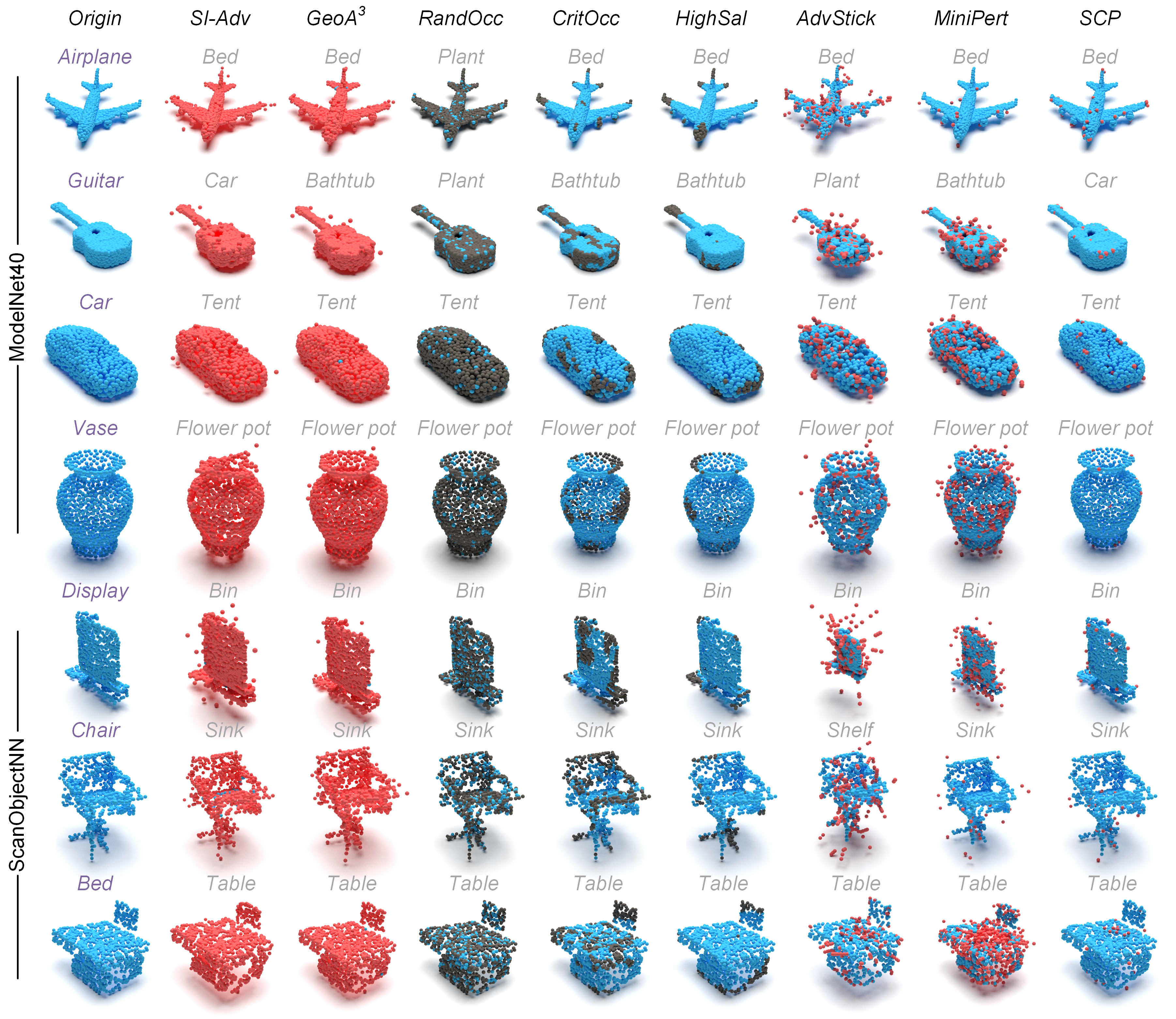}

    \caption{Visualization of original and adversarial point clouds generated by different attack methods targeting PointNet on ModelNet40 and ScanObjectNN. Blue points represent original points, red points denote  perturbed  points, and black points indicate removed points. Ground truth and predicted labels are shown above each point cloud in purple and gray, respectively.}
  \label{fig:vis_adv}

\end{figure*}

\firstpara{Comparison with Dense Attacks}  
To assess the effectiveness of using fewer but cooperative perturbations, we compare SCP with dense baselines that modify nearly all points. Tab.~\ref{tab:dense_impercep_metrics} summarizes the perturbation sizes required to achieve the highest ASR on ModelNet40 and ScanObjectNN across various classifiers. Dense attacks including PGD, IFGM, GeoA$^{3}$, AOF, and SI-Adv reach 100\% ASR while introducing high distortion, as shown by large CD, HD, and other imperceptibility metrics. In contrast, SCP also achieves 100\% ASR while modifying fewer than 50 points on average, which is two orders of magnitude fewer than dense methods. It also produces lower distortion on most metrics such as CD, HD, GR, Curv, and EMD. These results show that even when dense methods modify almost the entire input, our cooperative framework offers better imperceptibility with significantly fewer changes.

\if 0
\firstpara{Comparison with Dense Attacks}  
To further assess the effectiveness of using fewer but cooperative perturbations, we compare SCP with dense attack baselines that modify almost all points. Tab.~\ref{tab:dense_impercep_metrics} summarizes the perturbation sizes required by different methods to achieve their highest ASR on both ModelNet40 and ScanObjectNN across various classifiers. Dense attack baselines, including PGD, IFGM, GeoA$^{3}$, AOF, and SI-Adv, achieve 100\% ASR by modifying nearly all points but incur high distortion, as indicated by large CD, HD, and other imperceptibility metrics. In contrast, SCP attains the same perfect ASR while modifying fewer than 50 points on average, which is two orders of magnitude fewer than dense baselines, and achieves lower distortion on most evaluated metrics, including CD, HD, GR, Curv, and EMD, across both datasets. These results demonstrate that even when dense methods modify almost the entire input, our cooperative perturbation framework delivers imperceptibility that is in most cases superior, while requiring drastically fewer modifications.
\fi

\firstpara{Visualization}  
Fig.~\ref{fig:vis_adv} shows adversarial point clouds generated by various attack methods against PointNet. Most sparse approaches  remove points, creating visible holes, or perturb points with large displacements, both of which degrade geometric coherence. SCP, in contrast, perturbs only a small number of points with small magnitudes. Together with the quantitative results in Tab.~\ref{tab:sparseAttack}, which show that SCP achieves the highest attack success rates, e.g., 100\%, these observations confirm that SCP effectively balances attack strength and imperceptibility.

\begin{figure*}[!t]
    \centering
    \includegraphics[width=0.87\linewidth]{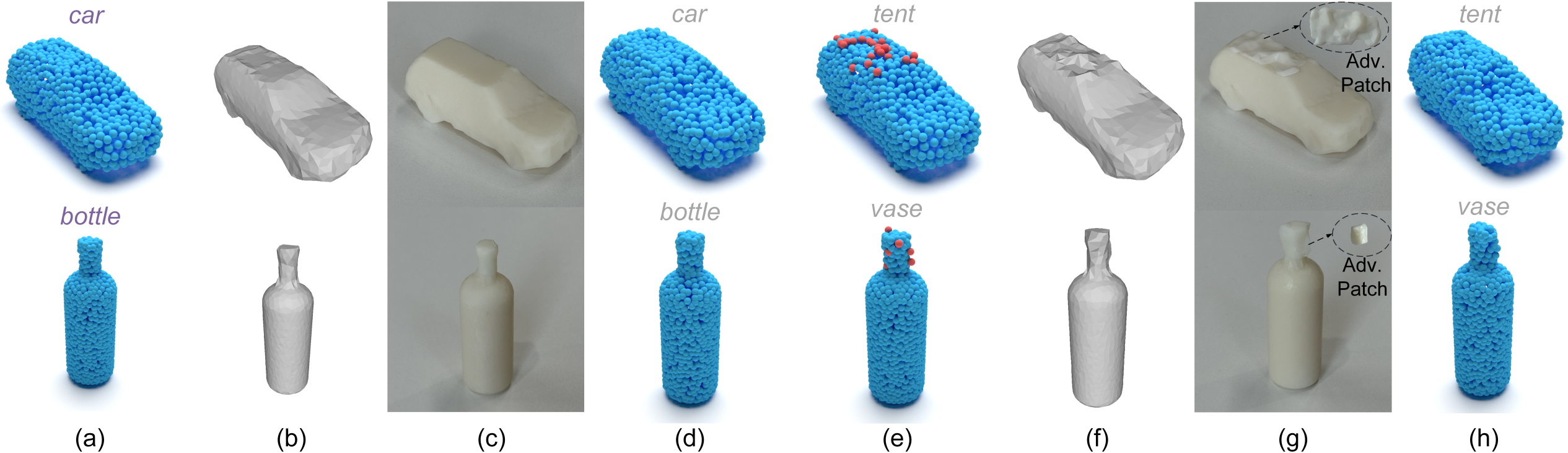}

\caption{  
Physical attack results of SCP.  
(a) Original point clouds, (b) reconstructed meshes, (c) 3D-printed objects,  
(d) re-scanned point clouds of (c),  
(e) adversarial point clouds generated by   SCP,  
(f) their reconstructed meshes,  
(g) physical adversarial objects obtained by attaching the 3D-printed patches to (c), and  
(h) re-scanned adversarial point clouds captured from the patched objects in (g).  
Purple text indicates ground-truth labels, while gray text indicates predictions by PointNet.  
%A spatial proximity constraint is enforced during cooperative subset expansion to ensure perturbations remain clustered for practical patch fabrication.  
}
  \label{fig:physical}
\end{figure*}

\begin{table*}[!t]
\centering

\setlength{\tabcolsep}{6pt}  % tighter column spacing
\scalebox{0.83}{
\begin{tabular}{rcccccccccccccc}
\toprule
\multirow{2}{*}{\# Points} &
  \multicolumn{7}{c}{{ModelNet40}} &
  \multicolumn{7}{c}{{ScanObjectNN}} \\
  \cmidrule(lr){2-8}\cmidrule(lr){9-15}
  & ASR & CD & HD & $l_2$ & GR & Curv & EMD & ASR & CD & HD & $l_2$ & GR & Curv & EMD \\
  & (\%) & ($10^{-4}$) & ($10^{-2}$) &  &  & ($10^{-2}$) & ($10^{-2}$) & (\%) & ($10^{-4}$) & ($10^{-2}$) &  &  & ($10^{-2}$) & ($10^{-2}$) \\
\midrule
2    &  87.792 & 0.878 & 6.060 & 0.449 & 0.270 & \textbf{0.040} & \textbf{0.062} &  89.305 & \textbf{0.852} & 4.662 & \textbf{0.211} & 0.131 & \textbf{0.031} & \textbf{0.031} \\
5    &  98.000 & 0.701 & 2.680 & \textbf{0.440} & 0.210 & 0.060 & 0.090 &  96.805 & 0.990 & 1.711 & 0.338 & 0.111 & 0.052 & 0.048 \\
10   &  98.708 & \textbf{0.607} & 1.920 & 0.443 & 0.167 & 0.097 & 0.125 &  99.400 & 1.074 & 0.645 & 0.362 & 0.100 & 0.077 & 0.070 \\
20   &  99.190 & 0.795 & 0.908 & 0.476 & 0.165 & 0.113 & 0.178 &  99.960 & 0.985 & 0.598 & 0.489 & 0.094 & 0.110 & 0.103 \\
$\sim$30   & \textbf{100} & 0.741 & \textbf{0.315} & 0.521 & \textbf{0.160} & 0.148 & 0.199 & \textbf{100} & 0.931 & \textbf{0.416} & 0.665 & \textbf{0.091} & 0.133 & 0.167 \\
\bottomrule
\end{tabular}
}
\caption{Impact of cooperative subset size on SCP performance against PointNet on ModelNet40 and ScanObjectNN.}
\label{tab:set_size}

\end{table*}

\begin{table}[!t]
\centering
\setlength{\tabcolsep}{1.5pt}

\scalebox{0.83}{%
\begin{tabular}{l l *{7}{c}}
\toprule
{Model} & {Hessian} &
ASR & CD & HD & $l_2$ & GR  & \# Points & Time \\
& & (\%) & ($10^{-4}$) & ($10^{-2}$) &  &  & &(s)\\
\midrule
\multirow{2}{*}{PointNet}
 & Full  & 100 & 0.746 & 0.297 & 0.509 & 0.155  & 34 & 87.89\\
 & Sparse        & 100 & 0.741 & 0.315 & 0.521 & 0.160  & 34 & 11.66\\
\midrule
\multirow{2}{*}{DGCNN}
 & Full  & 100 & 1.044 & 0.840 & 1.370 & 1.500  & 63 & 197.96\\
 & Sparse        & 100 & 1.059 & 0.841 & 1.378 & 0.150 & 62 & 24.27\\
\midrule
\multirow{2}{*}{PTv1}
 & Full  & 100 & 1.012 & 0.881 & 0.796 & 0.138  & 52 & 349.15\\
 & Sparse        & 100 & 1.010 & 0.877 & 0.795 & 0.139 & 52 & 43.47\\
\midrule
\multirow{2}{*}{Mamba3D}
 & Full  & 100 & 0.824 & 0.788 & 1.005 & 0.157  & 49 & 285.25\\
 & Sparse        & 100 & 0.831 & 0.794 & 1.010 & 0.158 & 49 & 37.09\\
\bottomrule
\end{tabular}}
\caption{Comparison of SCP with full Hessian-based (Full) and greedy (Sparse) subset selection for attacking four DNN models on ModelNet40.}
\label{tab:hessian_vs_ours}
\end{table}

\subsection{Ablation Studies and Other Analysis}

\firstpara{Effect of Cooperative Subset Size}
To assess the impact of cooperative subset size, we evaluate SCP under varying limits on the number of perturbed points, as shown in Tab.~\ref{tab:set_size}. The results reveal a clear trend: even with only two perturbed points, SCP attains ASR close to 90\% on both datasets. Increasing the subset size to 5–10 points significantly boosts ASR, while distortion metrics such as CD, HD, $l_2$, and EMD remain low. When the subset size exceeds 20 points, SCP achieves near-perfect ASR and further reduces HD,  indicating smaller maximum point displacements.
At around 30 points, SCP reaches 100\% ASR on both datasets with distortion still an order of magnitude lower than dense baselines.
These results confirm that cooperative selection allows SCP to achieve strong attacks with very few sparse perturbations, validating the efficiency of the proposed approach.

\firstpara{Effect of Different Cooperative Subset Selection Strategies}  
To assess whether our greedy selection compromises performance, we compare SCP using full Hessian-based subset selection with SCP using the proposed strategy. The full Hessian variant identifies cooperative points by directly computing and analyzing the complete Hessian matrix. As shown in Tab.~\ref{tab:hessian_vs_ours}, the greedy approach achieves almost the same attack success, distortion metrics, and number of selected cooperative points as the full Hessian method, while reducing computation time by an order of magnitude, confirming that it preserves effectiveness at much lower cost.

\firstpara{Physical  Validation}  
We further evaluate SCP in a physical setting by transforming adversarial point clouds into real-world objects through 3D printing. To enable patch fabrication, the cooperative subset expansion is intentionally constrained to select spatially close points, forming a localized perturbation region. This region is 3D-printed as a detachable patch and affixed to the clean object. After re-scanning and re-sampling the patched object, the resulting point cloud is fed back into the classifier. As shown in Fig.~\ref{fig:physical}, the clean samples are correctly classified, whereas the patched objects consistently cause misclassification, confirming that SCP successfully transfers to the physical domain.

\begin{figure}[!t]
\centering
    \includegraphics[width=0.8\linewidth]{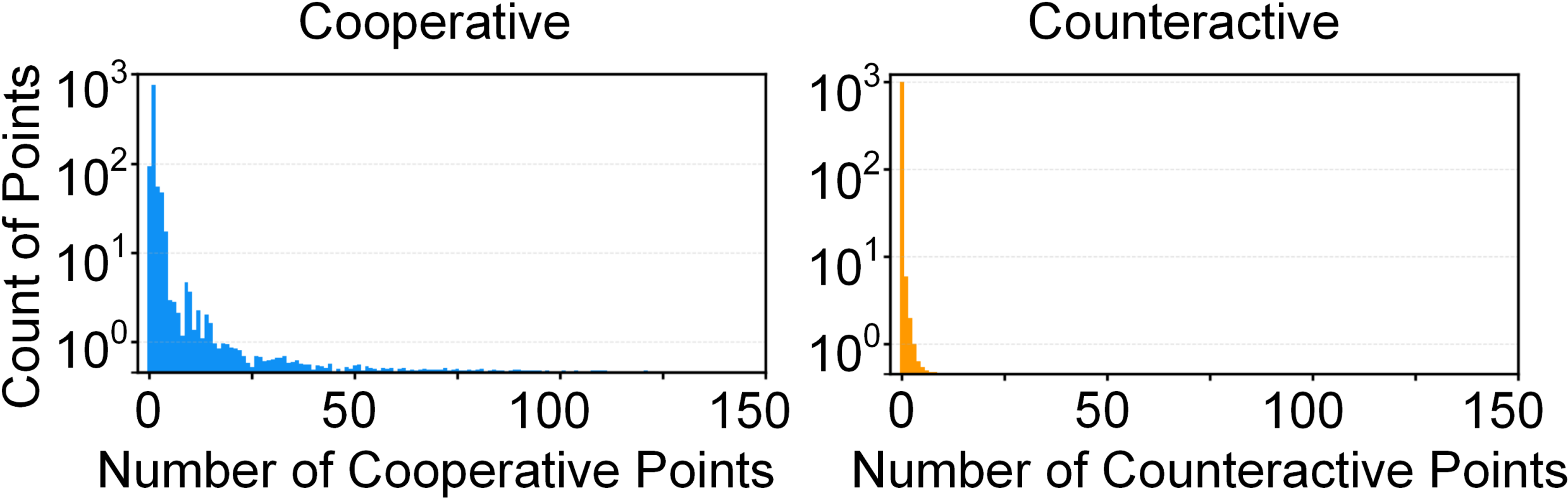}

\caption{Frequency distributions of cooperative and counteractive relationships for PointNet on ModelNet40.}
  \label{fig:Cooperative}

\end{figure}

\firstpara{Analyzing Cooperative/Counteractive Relationship}  
For each perturbed point, we calculate,  in a pairwise manner, the number of other points satisfying Eqn.~\ref{eq:cooperation} with ``$>$'' (cooperative) or ``$<$'' (counteractive), evaluated on ModelNet40 using PointNet. The distributions in Fig.~\ref{fig:Cooperative} show that cooperative counts are concentrated at small values but span a wide range, with a few points connected to dozens or even over a hundred others. In contrast, counteractive counts remain very low and drop sharply as the count increases, with only a negligible number of points showing higher values.

\section{Conclusion}

In this paper, we have presented SCP, a sparse and cooperative perturbation framework for adversarial attacks on 3D point clouds. SCP selects a compact yet cooperative subset of points, where restricted perturbations still amplify adversarial effects.
 Extensive experimental results show that SCP consistently achieves 100\% attack success rates while maintaining high imperceptibility, outperforming both state-of-the-art sparse and dense methods. In future work, we plan to extend SCP to black-box and transferable attack settings.

\section*{Acknowledgements}
%We thank the anonymous reviewers for their valuable comments.
This work was supported in part by the National Natural Science Foundation of China (62472117, 62572400, U2436208, 62372129), the Guangdong Basic and Applied Basic Research Foundation (2025A1515010157, 2024A1515012064), the Science and Technology Projects in Guangzhou (2025A03J0137, 2024B0101010002), 
the CCF-NetEase ThunderFire Innovation Research Funding (CCF-Netease 202514),
the Project of Guangdong Key Laboratory of Industrial Control System Security (2024B1212020010),
the Key Laboratory Project of Computing Power Network and Information Security, Ministry of Education (2023ZD02),
and the High-Quality Talent Training Program – Graduate Student Development Project of the Graduate School, Guangzhou University.

\bibliography{aaai2026}

\end{document}